\documentclass[
  reprint,
  aps,
  prd,
  superscriptaddress,
  nofootinbib,
  floatfix
]{revtex4-2}

\usepackage{graphicx}
\usepackage{amsmath}
\usepackage{color}
\usepackage{units}
\usepackage{appendix}
\usepackage{tikz}
\usepackage{hyperref}
\usepackage{lineno}

\newcommand{\SPA}{School of Physics and Astronomy, Monash University, Vic 3800, Australia}
\newcommand{\OzGravMonash}{OzGrav: The ARC Centre of Excellence for Gravitational Wave Discovery, Clayton VIC 3800, Australia}

\newcommand{\berkeley}{Department of Physics, University of California, Berkeley, Berkeley, CA 94720, USA}

\newcommand{\bilby}{\texttt{Bilby}}

\newcommand{\pbilby}{\texttt{pBilby}}

\begin{document}

\title{Licence to Bin: Accurate and Scalable Inference for Binary Neutron Stars in Next-Generation Gravitational-Wave Detectors}

\author{Nir Guttman}
\email{nir.guttman@monash.edu}
\affiliation{\SPA}
\affiliation{\OzGravMonash}

\author{A. Makai Baker}
\affiliation{\berkeley}

\author{Paul D. Lasky}
\affiliation{\SPA}
\affiliation{\OzGravMonash}

\author{Eric Thrane}
\affiliation{\SPA}
\affiliation{\OzGravMonash}

\begin{abstract}
Next-generation gravitational-wave observatories will observe binary neutron-star mergers with much higher signal-to-noise ratios, over much longer durations and across broader frequency bands than current detectors. These long-duration signals present a major computational challenge for Bayesian parameter estimation. Reduced-order quadrature is a promising approach for accelerating inference, but in this regime its standard construction encounters severe memory and accuracy limitations. We present a practical reduced-order quadrature construction for long binary neutron-star signals with time-dependent detector response and full effects of the observatories' free-spectral range. Our approach combines improved adaptive frequency sampling, disk-backed streaming, and subbanded reduced-order quadrature construction, enabling efficient and accurate reduced-order models for signals that were previously intractable within the reduced-order quadrature framework. We demonstrate for the first time reduced-order Bayesian inference on an approximately 2 h binary neutron-star signal extending down to 5 Hz and with signal-to-noise ratio 2090. We show the resulting reduced-order quadrature remains sufficiently accurate for practical inference. The full analysis is carried out in about 48\,h using 128 CPU cores. We also find that, when time-dependent detector-response effects are included, a single Cosmic Explorer detector can localize such a signal to a 90\% credible sky area of approximately $41 \mathrm{deg}^2$, with important implications for multimessenger astronomy and cosmology. These results demonstrate that reduced-order methods can make next-generation binary neutron-star inference computationally feasible.
\end{abstract}

\maketitle

\section{Introduction}
The next generation of ground-based gravitational-wave observatories, including Cosmic Explorer~\citep{CE} and the Einstein Telescope~\citep{ET}, will transform gravitational-wave astronomy by detecting binary neutron-star mergers with unprecedented signal-to-noise ratios (SNRs) and over broader observing bands than current instruments~\citep{LIGO,virgo,kagra}. Access to low frequencies will keep binary neutron-star signals in band for tens of minutes rather than seconds, enabling dramatically improved measurements of properties such as sky position, masses, and spins.

This scientific opportunity comes with a major computational challenge. Bayesian parameter estimation for gravitational-wave signals~\citep{PE_orig} is already expensive for current detectors, and becomes substantially more demanding for long-duration, high-SNR signals~\citep{NextGenCost}. Likelihood evaluation requires repeated waveform generation and repeated computation of noise-weighted inner products over a large frequency array, implying analyses can become prohibitively expensive. Considerable effort has therefore been devoted to accelerating parameter estimation~\citep{rom_original,relative_binning1,methods2,methods2_2,Hu_2025,multi_banding,methods5,methods5.1,methods6,methods1,methods7}.

More specifically, a recent studies sought to address long-duration, high-SNR binary-neutron-star signals in next-generation detectors, including time-dependent detector-response effects, using relative binning and multibanding~\citep{BNS_localization2, realtive_binning_bns}. Heterodyning-based methods such as relative binning rely on a fiducial reference waveform, and their accuracy is expected to be highest in the local region of parameter space where the analyzed waveform remains close to that reference~\citep{relative_binning1}. In Ref.~\citep{realtive_binning_bns}, the injected waveform itself served as the fiducial reference, an idealized choice that is not practical for realistic analyses where the source parameters are not a priori known. The same study also developed and validated a multibanding likelihood including frequency-dependent detector-response effects, but did not present full parameter-estimation results based on that approximation. Even so, such developments are important for establishing the feasibility of inference in this challenging regime while clarifying the limitations of the available approximation strategies.

Reduced-order methods aim to exploit redundancy in waveform families in order to accelerate likelihood evaluation while preserving accuracy. One first constructs a reduced basis that provides a compressed representation of the waveform family. Empirical interpolation is then used to identify a small set of special frequencies from which the waveform can be reconstructed accurately, thereby defining a reduced-order model (ROM). This ROM is subsequently used to construct a reduced-order quadrature (ROQ) rule that replaces the standard likelihood evaluation with a much smaller quadrature. Earlier work showed that, for compact binary coalescence signals observed by current detectors at modest SNRs, this framework can accelerate gravitational-wave inference by orders of magnitude~\citep{ROM_orig,PhysRevD.108.123040,Smith2016}.

For binary neutron stars in next-generation detectors, reduced-order methods therefore appeared to offer a promising route toward making long-duration inference tractable. Previous studies showed that ROM and ROQ constructions could provide substantial likelihood acceleration for long binary neutron-star signals. Within that framework, Bayesian inference on signals lasting up to roughly 90 minutes was shown to be computationally achievable~\citep{Smith2021}, suggesting that ROQ could provide a viable foundation for parameter estimation in the high-SNR regime.

However, more recent work has sharpened this picture and exposed deeper difficulties~\citep{Baker2025}. This study showed that when one fully incorporates the time-dependent detector response due to the rotation of the Earth, together with free-spectral-range effects~\citep{FSR,FSR2,FSR3} that become relevant for long-arm detectors such as Cosmic Explorer, the waveform manifold becomes substantially more complex. 
\citet{Baker2025} demonstrate that, while the construction of ROMs and ROQs remains feasible for relatively low-SNR binary neutron-star signals of modest duration, it becomes computationally prohibitive for long-duration, high-SNR signals. This arises because, for sufficiently long signals, the detector is no longer stationary. As the Earth rotates, the detector orientation changes during the observation, continuously modulating the measured signal, while free-spectral-range effects introduce additional frequency dependence in the detector response, primarily at high frequencies. These effects add structure that cannot be captured by an approximately
constant detector response,\footnote{An approximately constant detector response
is commonly assumed in analyses of current gravitational-wave detectors.}
making the waveform manifold more difficult to represent with a compact ROM. In this regime, two significant challenges arise. 

The first is a memory limitation. The size of the reduced basis required for unbiased inference grows to the point that reduced-order modeling becomes impractical with contemporary computational resources. Although such requirements may still be feasible on some modern systems, they remain a substantial practical constraint. In particular, Ref.~\cite{Baker2025} showed that a memory requirement of $\gtrsim 100$\,GB effectively prevents analyses below approximately 16\,Hz. Since for binary neutron stars, low frequencies account for most of the signal duration, this result points to a genuine bottleneck for astrophysical inference in next-generation observatories.

The second is a precision limitation. \citet{Smith2016} showed that the empirical interpolation step can degrade the final ROQ accuracy by roughly two orders of magnitude relative to the reduced-basis accuracy. In principle, however, such degradation may still be acceptable if the underlying basis is sufficiently accurate, even for very high SNRs. By contrast, \citet{Baker2025} reported that even when the reduced-basis construction tolerance is set at machine precision, the resulting ROQ accuracy is degraded enough to impose a maximum SNR above which the ROQ approximation is no longer reliable.

In this paper, we solve these two remaining obstacles to ROQ-based inference in the long-duration, high-SNR regime and demonstrate that full Bayesian inference is feasible for next-generation binary neutron-star signals. Our approach substantially reduces the memory burden, improves numerical stability, and enables the construction to be parallelized. As a result, accurate ROQs can be built for signals that include the expected time dependence and extend to lower frequencies than was previously achievable, while maintaining sufficient accuracy for high-SNR inference.

The rest of the paper is organized as follows. Section~\ref{sec:roq_method} briefly reviews the construction of a standard ROQ. Section~\ref{sec:enhanced_roq} describes our approach for overcoming the limitations discussed above. In Section~\ref{sec:infernce}, we demonstrate these ideas using a $\sim2$\,h binary neutron-star signal with SNR $\simeq 2090$. In this extreme regime, we recover the chirp mass with $\mathcal{O}(10^{-6})\,M_\odot$ precision and obtain a 90\% credible sky area of approximately $41\,\mathrm{deg}^2$ with a single Cosmic Explorer detector, while validating that the ROQ is sufficiently accurate for inference. Finally, Section~\ref{sec:conclusions} presents our conclusions.

\section{Method}\label{sec:roq_method}

Reduced-order modeling and reduced-order quadrature are well established in the literature~\citep{RB,rom_review}. In this section, we briefly review the construction of reduced-order models and reduced-order quadrature rules for gravitational-wave inference. We do not introduce any new formalism here; rather, we summarize the established framework that forms the basis for the modifications presented in Section~\ref{sec:enhanced_roq}. 

The construction proceeds in three steps: first, one builds a reduced basis from a training set of waveforms; second, empirical interpolation is used to identify a small set of special frequencies at which the waveform is evaluated in the subsequent step; third, this interpolated representation is substituted into the likelihood to obtain a reduced-order quadrature (ROQ) rule. We follow the general framework developed in~\citep{ROM_orig,Smith2016}. 

\subsection{Reduced basis and empirical interpolation}
\label{subsec:rom}

Let $h(f;\theta)$ denote a frequency-domain waveform depending on source parameters $\theta$, sampled on a frequency grid of size $L$. The aim of reduced-order modeling is to approximate this waveform family using a much smaller set of basis functions. To this end, one first constructs a reduced basis
\begin{equation}
\{e_i(f)\}_{i=1}^{N_{\rm basis}},
\end{equation}
where $N_{\rm basis}$ denotes the number of basis elements. This basis is chosen such that $N_{\rm basis} \ll L$ and its span accurately represents the waveform family over the parameter region of interest.

The basis is built from a training set of waveforms using a greedy algorithm. Starting from an empty basis, the algorithm iteratively projects the training-set waveforms onto the current basis and adds the waveform with the largest projection error. This procedure is repeated until the maximum representation error falls below a prescribed tolerance. An important feature of the reduced-basis construction is that it provides direct control over the maximum approximation error, which is essential when the compressed model is later used for inference.

A central difficulty in constructing the ROM for long signals is that the underlying frequency grid becomes extremely large. For binary neutron-star signals extending to low frequencies, the signal duration in band can be hours, and the corresponding frequency resolution must be very fine. As a result, the size of the frequency grid can become too large to store and manipulate efficiently in memory. To alleviate this problem, \citet{Smith2016} introduced dynamic frequency sampling. In this approach, the full frequency range is divided into a sequence of bands, and each band is assigned its own frequency resolution. The resolution in a given band is determined by the duration of the longest waveform contribution expected in that frequency interval, evaluated empirically from the lightest binary system under consideration. This duration is rounded up to the next power of two, and the frequency spacing is then taken as its inverse. In this way, the waveform remains sampled above the Nyquist rate in every band, while avoiding the cost of using a uniformly fine frequency resolution across the full analysis range.

Once the reduced basis has been constructed, the waveform can be represented using empirical interpolation. The empirical interpolation method selects a set of $N_{\rm basis}$ frequency nodes $\{F_J\}_{J=1}^{N_{\rm basis}}$, with one node associated with each basis element, such that the waveform can be reconstructed from its values at those nodes. 
The resulting empirical interpolant is
\begin{equation}
h_{\rm ROM}(f;\theta)
=
\sum_{J=1}^{N_{\rm basis}} h(F_J;\theta)\,B_J(f),
\end{equation}
where $h(F_J;\theta)$ is the waveform evaluated with parameters $\theta$ at the frequency nodes $F_J$. The interpolation functions $B_J(f)$ are constructed from the reduced basis. Writing the reduced basis as $\{e_i(f)\}$, the interpolation functions take the form
\begin{equation}
B_J(f)=\sum_{i=1}^{N_{\rm basis}} e_i(f)\,(V^{-1})_{iJ},
\end{equation}
where the interpolation matrix is defined by
\begin{equation}
V_{iJ}=e_i(F_J).
\end{equation}
The reduced-order model is therefore completely specified by the reduced basis and the interpolation nodes. Since the waveform is reconstructed from values at only $N_{\rm basis}$ special frequencies rather than all $L$ points in the original grid, this representation is highly compressed when $N_{\rm basis} \ll L$. In contrast to the reduced-basis construction, the empirical interpolation stage is carried out on the entire frequency grid. The selected basis vectors are therefore reconstructed on the full grid and orthonormalized before the interpolation nodes are computed.

\subsection{Reduced-order quadrature}
\label{subsec:roq}

The reduced-order model can be used to accelerate likelihood evaluation. For Gaussian noise, the standard frequency-domain natural log-likelihood, up to an additive constant, is
\begin{equation}\label{eq:log_likelihood}
\log \mathcal{L}
=
\langle d,h(\theta)\rangle
-\frac{1}{2}\langle h(\theta),h(\theta)\rangle
-\frac{1}{2}\langle d,d\rangle,
\end{equation}
where $d$ is the detector data and $\langle \cdot,\cdot\rangle$ denotes the noise-weighted inner product~\citep{Thrane_Talbot_2019}. Direct evaluation of these inner products is expensive because it requires repeated waveform evaluation and summation over the full frequency grid.

Substituting the empirical interpolant into the likelihood replaces the full sum by a much smaller quadrature rule involving only the interpolation nodes. For the linear term, one obtains
\begin{equation}\label{eq:linear}
\langle d,h(\theta)\rangle
\approx
\sum_{J=1}^{N_{\rm L}} \omega_J\, h(F_{{\rm L},J};\theta),
\end{equation}
where the weights $\omega_J$ depend only on the data. The basis functions and the detector noise power spectral density can therefore be precomputed prior to inference.

Similarly, the quadratic term can be approximated by
\begin{equation}\label{eq:quad}
\langle h(\theta),h(\theta)\rangle
\approx
\sum_{I=1}^{N_{\rm Q}} \psi_I\,\bigl|h(F_{{\rm Q},I};\theta)\bigr|^2,
\end{equation}
where $\psi_I$ are quadrature weights for the quadratic part of the likelihood. The prescription for computing these weights can be found in~\citet{Smith2021}.
In general, the linear and quadratic terms may require different reduced bases and therefore different sets of interpolation nodes.

Substituting Eqs.~\ref{eq:linear} and~\ref{eq:quad} into Eq.~\ref{eq:log_likelihood}, we define the reduced-order quadrature likelihood approximation,
\begin{equation}\label{eq:roq_log_likelihood}
\begin{split}
\log \mathcal{L}_{\rm ROQ}
={}&
\sum_{J=1}^{N_{\rm L}} \omega_J\, h(F_{{\rm L},J};\theta) \\
&-\frac{1}{2}\sum_{I=1}^{N_{\rm Q}} \psi_I\,\bigl|h(F_{{\rm Q},I};\theta)\bigr|^2
-\frac{1}{2}\langle d,d\rangle .
\end{split}
\end{equation}
Since the approximations in Eqs.~\ref{eq:linear} and~\ref{eq:quad} propagate into Eq.~\ref{eq:roq_log_likelihood}, the resulting ROQ likelihood must be validated against the standard likelihood in Eq.~\ref{eq:log_likelihood}. For a waveform parameter point $\theta$, we therefore define the induced ROQ log-likelihood error as
\begin{equation}\label{eq:roq_error}
\Delta \log \mathcal{L}_{\rm ROQ}(\theta)
\equiv
\left|\log \mathcal{L}_{\rm ROQ}(\theta)-\log \mathcal{L}(\theta)\right|.
\end{equation}
This quantity is evaluated over a set of test waveform parameters and summarized using its distribution across that set.

The ROQ computational advantage arises because the cost of likelihood evaluation is reduced from a sum over the full frequency grid to a sum over only $N_{\rm L}$ and $N_{\rm Q}$ interpolation nodes, with typically $N_{\rm L},N_{\rm Q} \ll L$. In this way, ROQ accelerates the likelihood evaluation by exploiting the low-dimensional structure of the waveform family while preserving the accuracy of the calculation.

\section{An enhanced construction strategy for ROM and ROQ}\label{sec:enhanced_roq}

In this section, we revisit the workflow for constructing ROMs and ROQs introduced in~\citep{ROM_orig,Smith2016} in the regime of long-duration signals with time-dependent detector response. Our aim is not to alter the underlying formalism, but rather to modify the construction strategy so that it remains both computationally tractable and sufficiently accurate. To this end, we revise the dynamic-sampling procedure, introduce streaming and disk-backed memory management, and divide the full frequency range into subbands in order to control both memory usage and signal complexity. Finally, we define a practical procedure for ROQ construction that changes the way in which the resulting ROQ accuracy is assessed.
These developments are implemented in our software package \href{https://github.com/NirGutt/gwRombusX}{\texttt{gwRombusX}}, which extends \texttt{gwRombus}~\citep{git_makai}. The latter is built on the \texttt{Rombus} framework~\citep{git_smith}, which provides the software foundation for the modifications described below. We make \texttt{gwRombusX} publicly available, together with example scripts and \bilby\ \cite{bilby_paper,bilby_add} integration. Additional implementation details are given in Appendix~\ref{appx:rombus}.

We demonstrate that these modifications enable the construction of ROQs for significantly more complex gravitational-wave signals than was previously practical.

\subsection{Accuracy limitation}\label{subsec:accuracy_limitation}

As noted in Section~\ref{subsec:rom}, adaptive frequency sampling was introduced in~\citet{Smith2016} (their Sec.~III.D) as a way to reduce the memory footprint of ROM construction. For very long signals, such a strategy becomes even more important. However, we find that the number of frequency grid points used during the initial ROM construction directly affects the accuracy of the final ROQ approximation.

This can be understood as follows: if the initial reduced-basis construction is performed on a frequency grid that is too coarse, waveform structure that would otherwise require additional basis elements may not be resolved. When the basis is subsequently reconstructed on the full grid, this missing structure manifests itself as a loss of accuracy in the final ROQ approximation. This becomes especially important in the high-SNR regime, where the tolerance on ROQ errors is particularly stringent.

We therefore propose an alternative adaptive-sampling procedure that is more closely tailored to compact binary coalescence signals. We find that dividing the frequency range uniformly in the variable $f^{-5/6}$ yields the optimal choice of frequency bands for minimizing the total number of frequency grid points. At leading post-Newtonian order, this follows from the inspiral-time scaling $t(f)\propto f^{-8/3}$, which implies that such a banding distributes the accumulated waveform duration, and hence the required frequency resolution more evenly across the frequency range. The derivation is presented in Appendix~\ref{appx:binning_derivation}. For each band, we compute the time spent in that interval, using the chirp-time estimate for the lightest system under consideration, and define an effective duration $\Delta t_{\rm eff}$. The frequency resolution in the band is then set to
\begin{equation}\label{eq:enhancement_factor}
\Delta f = \frac{1}{2^{N}\,\Delta t_{\rm eff}},
\end{equation}
where the integer $N$, referred to here as the enhancement factor, controls how finely the band is sampled. Following~\citet{Baker2025}, we impose a maximum bin size of 1\,Hz.

\begin{figure}
    \centering
    \includegraphics[width=0.99\linewidth]{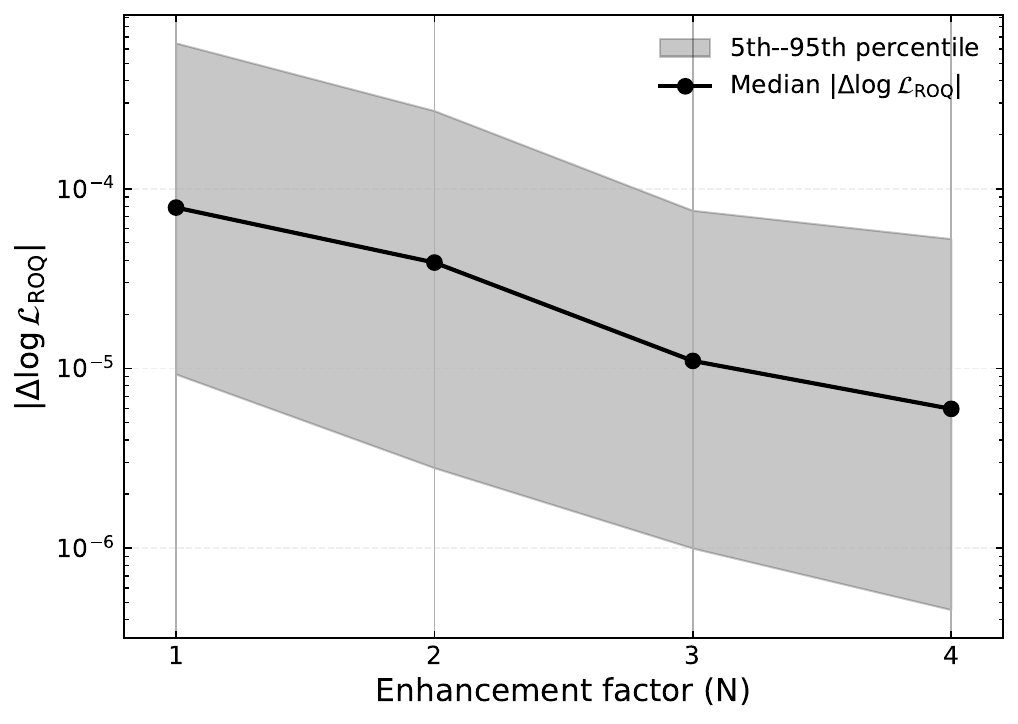}
    \caption{
    The ROQ log-likelihood error (Eq.~\ref{eq:roq_error}) as a function of the enhancement factor, $N$ (Eq.~\ref{eq:enhancement_factor}), used in the adaptive-sampling procedure, for an ROQ constructed over the frequency range 20--200\,Hz with a signal duration of 128\,s. The solid black curve shows the median and the shaded region indicates the 5th--95th percentile range over the tested waveform parameters. No signal is injected in this study. The figure illustrates that increasing the enhancement factor systematically improves the accuracy of the final ROQ.
    }
    \label{fig:accuracy_limitation}
\end{figure}

Figure~\ref{fig:accuracy_limitation} illustrates how the error decreases as the enhancement factor is increased for a representative construction over the frequency range 20--200\,Hz with a signal duration of 128\,s.

The value of the enhancement factor, $N$, is not unique, and must instead be determined by the required target accuracy. In Section~\ref{sec:roq_building}, we describe how this choice is made in practice. Thus, the dynamic sampling strategy must be treated as part of the accuracy budget of the ROQ construction, rather than only as a tool for reducing memory usage.

\subsection{Memory limitations}\label{subsec:memory_limitation}

The memory challenges identified by~\citet{Baker2025} stem from two main sources. Long-duration signals lead to extremely fine frequency spacing in the Fourier domain and are therefore represented by very large arrays across a broad band. In addition, including time-dependent detector-response effects leads to a dramatic increase in the complexity of the waveform manifold and hence the number of basis elements needed for an accurate representation. As a result, the training sets, bases, and intermediate matrices can become prohibitively large to store and manipulate in memory. 
To overcome this limitation, we adopt three complementary strategies. These considerations apply to the construction of both the ROM and the ROQ.

First, we replace the high in-memory workload with disk-backed streaming of the data. In practice, this means that large arrays generated during the construction are stored on disk and accessed in blocks, rather than being kept in memory all at once. Examples include the training set, the reduced basis, and intermediate matrices required for orthogonalization and empirical interpolation. This software-architecture solution required substantial modifications to the \texttt{gwRombus} code and allows us to control memory usage during the construction of both the ROM and the ROQ. In particular, we introduce a user-controlled memory cap that specifies the maximum amount of memory available to the construction procedure Appendix~\ref{app:disk_streaming} provides further details on the disk-backed implementation.

Second, as discussed in Section~\ref{subsec:accuracy_limitation}, we improve the selection of the basis frequencies. This substantially reduces the number of frequency grid points while keeping the selected frequencies close to optimal, thereby further lowering the memory requirements.

Third, we divide the full frequency range into subbands and construct a separate ROQ in each subband. This is justified by the standard assumption of stationary Gaussian noise, under which the likelihood factorizes across frequency bins~\citep{Thrane_Talbot_2019}. Because each subband ROQ is smaller than one constructed over the full frequency range, this immediately reduces the memory burden during construction. Moreover, different frequency regions generally require different numbers of basis elements, as we show in Section~\ref{sec:infernce}. Treating the full band simultaneously forces all regions to inherit the complexity of the most demanding one. This is especially problematic for very long signals, where the number of frequency grid points can reach into the millions.

More generally, subbanding allows the accuracy of each region to be controlled independently while maintaining a fixed total error budget. In turn, this makes it possible to relax the accuracy requirements in low-SNR regions (see Section~\ref{sec:roq_accuracy}) and thereby further reduce both memory usage and computational cost.

A concrete example of these benefits appear in our implementation, where the reduced basis on the full frequency grid is constructed using a Gram--Schmidt orthogonalization procedure (see Appendix~\ref{appx:rombus} for further implementation details). In this case, the main computational bottleneck arises from the fact that each candidate vector must be orthogonalized against all previously selected basis vectors, so that the cost grows approximately as $N_{\rm basis}^2$, where $N_{\rm basis}$ is the number of basis elements. For long-duration signals, each such operation must be carried out on extremely long vectors, making the construction prohibitively expensive. For example, a signal of duration 7500\,s corresponds to vectors with approximately $1.5\times10^7$ elements, requiring about 200\,MB of memory per vector. By contrast, these enhancements reduce the effective vector length in each construction, lower the cost of each orthogonalization step, and decrease the number of basis elements required in any one band. In addition, the subbands can be processed independently, making the construction naturally parallelizable and substantially faster, while keeping the memory requirements within the capabilities of standard computing hardware.

Although Gram--Schmidt is not essential to the ROQ construction, and alternative algorithms may offer improved numerical stability for very long vectors, we expect that any such approach would similarly benefit from reduced vector sizes, both in terms of memory usage and computational cost.

Taken together, these modifications provide a practical framework for controlling memory usage throughout the construction, and form the basis of the enhanced ROM and ROQ pipeline described in the remainder of this section.

\subsection{Assessing the ROQ accuracy}\label{sec:roq_accuracy}

Although a tolerance is imposed during the construction of the ROM, the subsequent construction of the ROQ can introduce a larger approximation error~\citep{Smith2016}. It is therefore necessary to assess the final ROQ accuracy through the log-likelihood error defined in Eq.~\ref{eq:roq_error}. Following~\citet{relative_log_likelihood2,PhysRevD.108.123040,Baker2025}, one may impose the standard requirement
\begin{equation}\label{eq:SNR_accuracy}
\frac{\Delta \log \mathcal{L}_{\text{ROQ}}}{\log \mathcal{L}} < \frac{1}{\rho^2},
\end{equation}
where $\rho$ is the SNR. This criterion follows from the condition that modeling errors remain smaller than the statistical uncertainty of a signal with SNR $\rho$, so that the ROQ approximation does not introduce appreciable bias into the inference.

However, this threshold is overly conservative. Far from the likelihood peak, where $\log \mathcal{L}$ decreases rapidly, the relative error can become artificially large in regions of little relevance for inference. We therefore compare $\Delta \log \mathcal{L}_{\text{ROQ}}$ to a characteristic scale set by the expected distribution of the log-likelihood. Using the result derived in Appendix~\ref{app:logL_error}, the log-likelihood near the peak has mean $\frac{1}{2}\rho_{\mathrm{opt}}^2$ and fluctuations of order $\rho_{\mathrm{opt}}$.

Motivated by this scaling, we define the reference quantity
\begin{equation}
\log \mathcal{L}_{\rm ref} \sim \mathcal{N}\!\left(\frac{1}{2}\rho_{\mathrm{opt}}^2,\rho_{\mathrm{opt}}^2\right).
\end{equation}
In practice, for each evaluation of $\Delta \log \mathcal{L}_{\text{ROQ}}$, we draw samples from this reference distribution and compute the normalized-error samples
\begin{equation}
r_j = \frac{\Delta \log \mathcal{L}_{\text{ROQ}}}{\log \mathcal{L}_{{\rm ref},j}},
\end{equation}
where $j$ labels the samples drawn from the reference distribution.
We then take the 95th percentile of the resulting distribution as our summary statistic,
\begin{equation}
r_{95} \equiv {\rm perc}_{95}\!\left(\left\{r_j\right\}\right).
\end{equation}
Following Eq.~\ref{eq:SNR_accuracy}, we require
\begin{equation}\label{eq:95_critiria}
r_{95} < \rho^{-2},
\end{equation}
which ensures that the ROQ likelihood error remains below the target accuracy at the relevant SNR.

This avoids overemphasizing errors in regions where the likelihood is already negligible, while retaining the correct SNR scaling near the peak. 
Although applying the criterion of Eq.~\ref{eq:95_critiria} requires a prior estimate of the target SNR, this does not present a practical difficulty, since an approximate SNR is already available from the matched-filter detection pipeline~\citep{SNR_detection_pipeline}. More generally, the SNR could be estimated iteratively: starting from a fiducial value, one could construct an ROQ, assess its adequacy after an initial inference run, and then tighten the ROQ requirements and rebuild if needed.
Thus, $r_{95}$ provides a practical, SNR-aware criterion for determining whether a given ROQ construction is sufficiently accurate for inference.

\subsection{Distributing the ROQ error budget}\label{sec:roq_error_bugdet}

Once the full frequency range is divided into subbands, the total allowed ROQ error must be distributed among them. We denote the total allowed error budget by $\epsilon_{\rm tot} \equiv \rho^{-2},$
following Eq.~\ref{eq:95_critiria}. Although the total error budget, $\epsilon_{\rm tot}$, is fixed by the target SNR, its allocation across subbands is not unique.

In general, the computational cost of constructing the ROQ in a given band depends on the number of frequency grid points in that band, the SNR, and the target accuracy assigned to that band. It may also depend on prior expectations about the modeling complexity in different frequency regions. For example, in binary neutron-star signals, the low-frequency regime is expected to be more expensive to model because it contains long-duration time-dependent effects due to Earth's rotation. In practice, such effects complexity can be partially absorbed into the choice of subbands. 

For simplicity, however, we adopt the following model.
Let $\epsilon_i$ denote the ROQ error budget assigned to subband $i$, let $N_i$ denote the number of frequency grid points in that band, and let $\rho_i$ denote the corresponding contribution to the total SNR. We write the total construction cost as
\begin{equation}\label{eq:rom_cost}
C \propto \sum_i \frac{N_i^{\alpha}\rho_i^{\beta}}{\epsilon_i^{\gamma}},
\end{equation}
where $\alpha$, $\beta$, and $\gamma$ are positive constants.  This expression is intended only as a leading-order proxy for how the cost scales with subband size, signal strength, and target accuracy; it is not meant to represent a unique or exact model of the construction cost. The subband errors are constrained by the total error budget,
\begin{equation}
\sum_i \epsilon_i^2 = \epsilon_{\rm tot}^2.
\end{equation}
Minimizing the cost under this constraint, and setting $\alpha=\beta=\gamma=1$ yields\footnote{The specific choice of cost model is heuristic and is adopted for simplicity. It provides a simple baseline for distributing the error budget across subbands.}
\begin{equation}
\epsilon_i \propto (N_i\rho_i)^{1/3}.
\end{equation}
with the proportionality constant fixed by the total-error constraint. The derivation is given in Appendix~\ref{appx:error_budget}.

This prescription assigns a larger error allowance to subbands that are expected to be more expensive to model, while preserving the required total error budget. For each subband, we therefore impose
\begin{equation}
r_{95,i} < \epsilon_i,
\end{equation}
where $r_{95,i}$ is the 95th percentile of the summary statistic in subband $i$. If this condition is satisfied in every subband, the combined ROQ error remains consistent with the total-accuracy requirement in Eq.~\ref{eq:95_critiria}.

\subsection{Construction of the ROQ}\label{sec:roq_building}

Using the subbanding, adaptive-sampling, and memory-management strategies described above, we construct the ROQ iteratively and independently within each frequency subband. The procedure is as follows:

\begin{enumerate}
    \item Divide the full frequency range into subbands.
    \item For each subband, specify a target ROQ accuracy consistent with the total error budget, following Section~\ref{sec:roq_error_bugdet}.
    \item Choose the adaptive frequency sampling in that subband, i.e., the number of frequency grid points or, equivalently, the enhancement factor $N$ introduced in Section~\ref{subsec:accuracy_limitation}.
    \item Construct the ROM and ROQ for that subband.
    \item Build the corresponding likelihood weights and assess the ROQ accuracy using the criterion described in Section~\ref{sec:roq_accuracy}.
    \item If the target accuracy is not met, increase the sampling density in that subband (step 3) and repeat steps 4--5 until the required accuracy is achieved.
\end{enumerate}

This procedure allows the sampling density, basis size, and final ROQ accuracy to be controlled independently in each frequency region. In particular, it makes it possible to allocate finer sampling only where demanded by the modeling complexity, while keeping the overall construction computationally tractable.

\section{Inference with a high-SNR binary neutron-star signal}\label{sec:infernce}
To demonstrate the performance of the method, we carry out Bayesian inference on a representative long-duration, high-SNR binary neutron-star signal observed in a single Cosmic Explorer detector~\cite{CE_sensetivity}. We adopt a single-detector setup in order to illustrate how the time-dependent detector response, modulated by the Earth’s rotation, can improve sky localization. We first describe the inference setup and the corresponding ROQ construction, and then assess the quality of the approximation through likelihood reweighting.

\subsection{Inference setup}\label{sec:inference_setup}

We consider a binary neutron-star signal with minimum frequency $f_{\min}=5\,\mathrm{Hz}$, maximum frequency $f_{\max}=2048\,\mathrm{Hz}$, analyzed within a data stretch of duration $T=7500\,\mathrm{s}$. The time-dependent interferometer response includes both the Earth's rotation and free-spectral-range effects.\footnote{The implementation is described in~\citet{Baker2025} and is based on the associated \bilby \;development documented in \url{https://git.ligo.org/lscsoft/bilby/-/merge_requests/1370}.}

For the waveform model we use \texttt{IMRPhenomPv2\_NRTidal}~\citep{NRTidal,pv2a}. We adopt standard binary-neutron-star priors~\citep{gw170817,bilby_paper} on the full 17-dimensional parameter space, with the exception that we impose a restricted prior on the chirp mass, following the strategy of~\citet{Baker2025,Smith2021}. This choice is motivated by the fact that the chirp mass is already well constrained by the detection pipeline, and restricting the prior to a narrow region around the target value substantially reduces the cost of the ROQ construction while remaining well motivated. In addition, \citet{Smith2021} notes that, although narrow chirp-mass patches are convenient for individual ROM constructions, broader priors can in principle be covered by combining multiple local bases.
For the tidal deformability parameters, we impose the physical constraint \(\Lambda_2 \ge \Lambda_1\), assuming the standard mass ordering \(m_1 \ge m_2\)~\citep{Tidal_deformability}.
The full prior choices, ranges, and injection values used in this study are listed in Appendix~\ref{appx:inference}.

We draw the source parameters randomly from the adopted prior, with the exception of the luminosity distance, which we set manually to $20\,\mathrm{Mpc}$ in order to obtain an SNR of approximately $\rho \simeq 2090$ in a single detector. We choose this distance deliberately in order to stress test the ROQ construction in a particularly demanding high-SNR regime. This is comparable to the network SNR expected in a two-detector observation of a GW170817-like event, in which each detector measures $\rho \sim 1500$~\citep{gw170817}.

We analyze a zero-noise realization of stationary Gaussian noise~\citep{zero_noise}. Statistically, this corresponds to the most likely noise realization, and when combined with sufficiently regular priors, the posterior peaks at the injected parameter values. This provides a useful diagnostic check of both the likelihood implementation and the resulting ROQ. We emphasize, however, that this need not hold exactly in the presence of parameter degeneracies: if different parameter combinations yield comparable likelihood values, the posterior can remain broad or elongated along the degenerate directions, and the injected values need not coincide precisely with the posterior peak. See Appendix~\ref{app:posterior_shift_diagnostic} for further details.

We use distance and phase marginalization. In \bilby{}~\citep{bilby_paper}, the distance marginalization is implemented using a precomputed lookup table, following the standard likelihood formalism described by~\citet{Thrane_Talbot_2019}. In the current \bilby \;implementation, this lookup table is constructed using a number of grid points that is suitable for the expected signals of current gravitational-wave detectors~\citep{gwtc3_detector}. 
However, find that this is insufficient for the high-SNR regime explored by Cosmic Explorer. In this regime, the likelihood is extremely sharply peaked, so a broad prior together with a fixed lookup-table resolution leads to an insufficiently accurate estimate of the marginalization integral. We therefore adopt a more finely sampled lookup table and place the grid in \emph{logarithmic} distance, which more efficiently resolves the high-SNR regime because the SNR scales approximately as the inverse of the luminosity distance, $d_L^{-1}$. This ensures a more accurate numerical evaluation of the marginalized integral.

Finally, we perform the inference using nested sampling~\citep{skilling2004,skilling2006} with \texttt{dynesty}~\citep{dynesty_paper}. Nested sampling for high-SNR signals is computationally demanding. We therefore employ \pbilby{} \citep{pbilby_paper}, running on 128 CPU cores with 4000 live points using the \texttt{acceptance-walk} stepping method with \texttt{naccept}=60. For this extremely challenging inference problem, the combined use of \pbilby, distance and phase marginalization, and ROQ is essential, since direct likelihood evaluation would otherwise be computationally prohibitive.\footnote{The modified versions of \bilby \;and \pbilby \;used in this work are available at \url{https://github.com/NirGutt/gwRombusX}}

\subsection{ROQ construction for the inference run}

The ROQ used in this analysis is constructed by dividing the full frequency range into a sequence of subbands, each with its own sampling density, basis size, and target accuracy.

In the present case, the low-frequency range is naturally separated into three main regimes: 5--10\,Hz, 10--30\,Hz, and 30--100\,Hz. In the first of these, the signal spends the largest amount of time in band, and the effect of the Earth's rotation is therefore most pronounced. In the second, the signal remains in band for tens of minutes, so time-dependent detector-response effects continue to play an important role. In the third, the impact of the Earth's rotation is substantially reduced. The remaining higher-frequency subbands are listed in Table~\ref{tab:subband_roq_summary} (see Appendix~\ref{appx:inference}). Although this subdivision is not unique, the number of basis elements ($N_{\text{basis}}$) required in each subband provides a useful diagnostic of both the local model complexity and the effective signal content in that frequency range.

For each subband, we divide the frequency range uniformly in $f^{-5/6}$ into 100 bins (see Section~\ref{subsec:accuracy_limitation} and Appendix~\ref{appx:binning_derivation}), thereby defining an initial frequency grid. We then adjust the enhancement factor until the resulting ROQ satisfies the accuracy criteria described in Sections~\ref{sec:roq_accuracy} and~\ref{sec:roq_error_bugdet}.

For the initial reduced-basis training set, we follow the prescription of~\citet{Baker2025}; however, we use $N_{\text{training}} \approx 4\times10^{5}$ training waveforms, about an order of magnitude fewer than in that work. This choice is intentional, and is meant to emphasize that when $N_{\text{basis}} \ll N_{\text{training}}$, the training set does not need to be extremely large. In the present case, the most complex subband requires $N_{\text{basis}} \sim 4000$, which is roughly two orders of magnitude smaller than the initial training set. This illustrates that, once the training set is comfortably larger than the final basis size, further enlarging it does not necessarily improve the practical quality of the construction.

For each ROQ construction, the reduced-basis tolerance is set to $10^{-15}$. We make this choice for two reasons. First, we find that at $10^{-16}$, i.e., near machine precision, the algorithm can stagnate and produce artificially large bases. Second, as pointed out by~\citet{Baker2025}, error estimates become unreliable in this regime. We therefore choose a slightly looser tolerance, while also doing so intentionally in order to demonstrate that the reduced-basis tolerance is not the only parameter controlling the final ROQ accuracy.

\subsection{Inference validation}

We validate the inference in two ways. First, because the analysis is carried out in a zero-noise realization, we expect the posterior to peak close to the injected parameter values, as discussed in Section~\ref{sec:inference_setup}. Second, we employ likelihood reweighting using the ROQ posterior as a proposal distribution and the standard gravitational-wave likelihood posterior as the target distribution, following the standard importance-sampling approach of~\citet{reweight}.

For posterior samples $\{\theta_i\}$ drawn from the ROQ analysis, the unnormalized importance weights are defined by
\begin{equation}
w_i \propto \frac{\mathcal{L}(d|\theta_i)}
{\mathcal{L}_{\rm ROQ}(d|\theta_i)} .
\end{equation}
The reweighted posterior is then obtained by assigning normalized weights
$\hat w_i = w_i/\sum_j w_j$
to the ROQ samples. The corresponding evidence under the standard likelihood can be estimated from the same weights as
\begin{equation}
Z \approx Z_{\rm ROQ}\,\frac{1}{N}\sum_{i=1}^{N} w_i ,
\end{equation}
so that the shift in log evidence is
\begin{equation}
\Delta \log Z \equiv \log Z-\log Z_{\rm ROQ}
\approx \log\!\left(\frac{1}{N}\sum_{i=1}^{N} w_i\right).
\end{equation}

Finally, we quantify the reweighting efficiency through the effective sample size,
\begin{equation}
N_{\rm eff} = \frac{\left(\sum_i w_i\right)^2}{\sum_i w_i^2},
\end{equation}
and report the efficiency as
\begin{equation}
\epsilon = \frac{N_{\rm eff}}{N}.
\end{equation}
If the ROQ approximation is accurate, we expect both a negligible shift in the log evidence and a reweighting efficiency close to unity.

\subsection{Inference results}\label{sec:infernce_results}

\begin{figure}
    \centering
    \includegraphics[width=\linewidth]{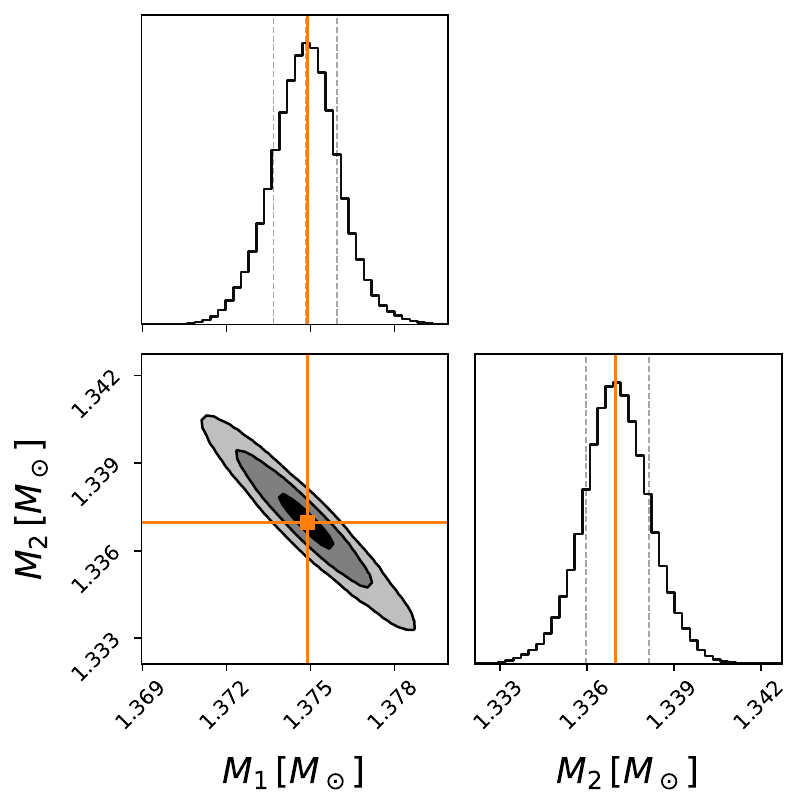}
        \caption{
            Marginalized posterior distributions for the source-frame component masses \(M_1\) and \(M_2\) of the injected binary neutron-star signal. In the one-dimensional panels, the solid orange lines indicate the injected values and the dashed lines show the \(\pm 1\sigma\) credible interval about the posterior median. The posteriors peak close to the injected parameters, as expected for the zero-noise realization adopted in this study.
            }
    \label{fig:corner_main_mass}
\end{figure}

Figure~\ref{fig:corner_main_mass} shows the marginalized posterior distributions for the component masses. The posterior peaks close to the injected parameter values. This provides a first consistency check that the ROQ likelihood reproduces the structure of the standard likelihood in the region of highest posterior support.

For these intrinsic parameters, we find that the component masses are measured with very high precision, with a 90\% credible interval width of approximately \(\Delta M_{1,2}^{90}\sim 4\times10^{-3}\,M_\odot\).

\begin{figure}
    \centering
    \includegraphics[width=\linewidth]{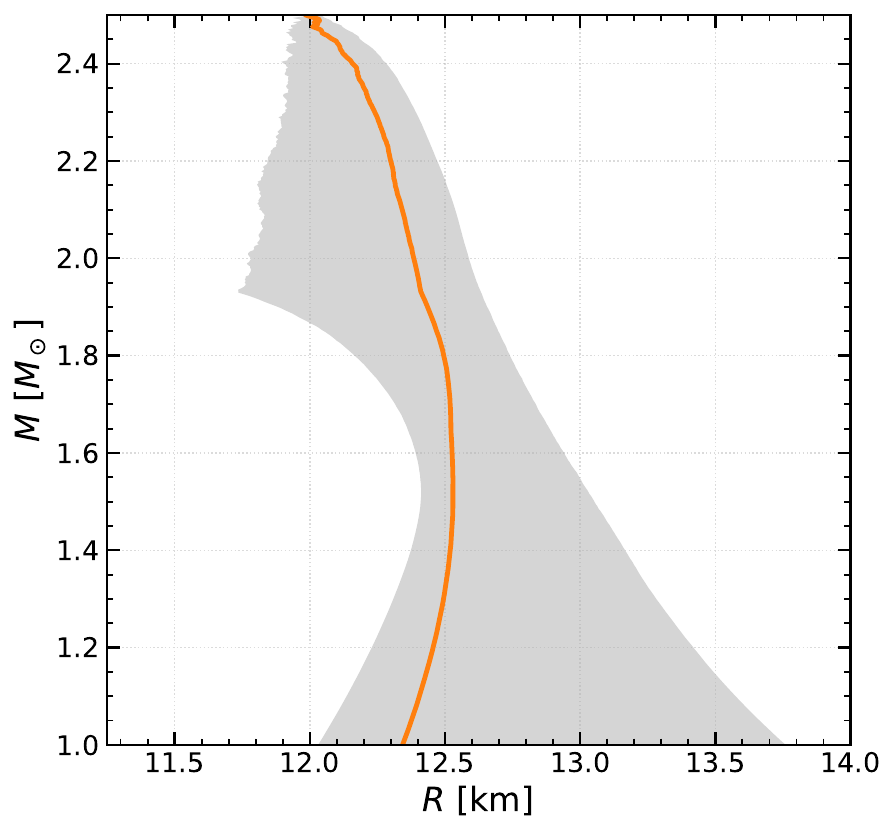}
    \caption{
    Mass--radius relation implied by the posterior on the polytropic equation of state. The orange curve denotes the median relation, and the gray band shows the 90\% credible region.
    }
    \label{fig:eos_mr}
\end{figure}

Figure~\ref{fig:eos_mr} shows the mass--radius relation implied by the posterior samples, obtained by inferring a polytropic equation of state with the \texttt{eosinference} framework~\citep{eosinference_code,eosinference}. The resulting relation is centered around \(R\sim 12.2\,\mathrm{km}\). 
In particular, for \(M \gtrsim 1.3\,M_\odot\), the width of the joint 90\% credible region in the mass–radius plane is below \(1\,\mathrm{km}\), reaching a minimum of approximately \(0.5\,\mathrm{km}\) near \(M\sim 1.6\)–\(1.7\,M_\odot\).
The joint mass–radius credible band therefore has sub-kilometer width over much of the astrophysically relevant neutron-star mass range. The posterior distributions for the tidal deformability parameters are provided in Appendix~\ref{appx:inference}.

\begin{figure}
    \centering
    \begin{tikzpicture}
        \node[anchor=south west, inner sep=0] (main) at (0,0)
            {\includegraphics[width=\linewidth]{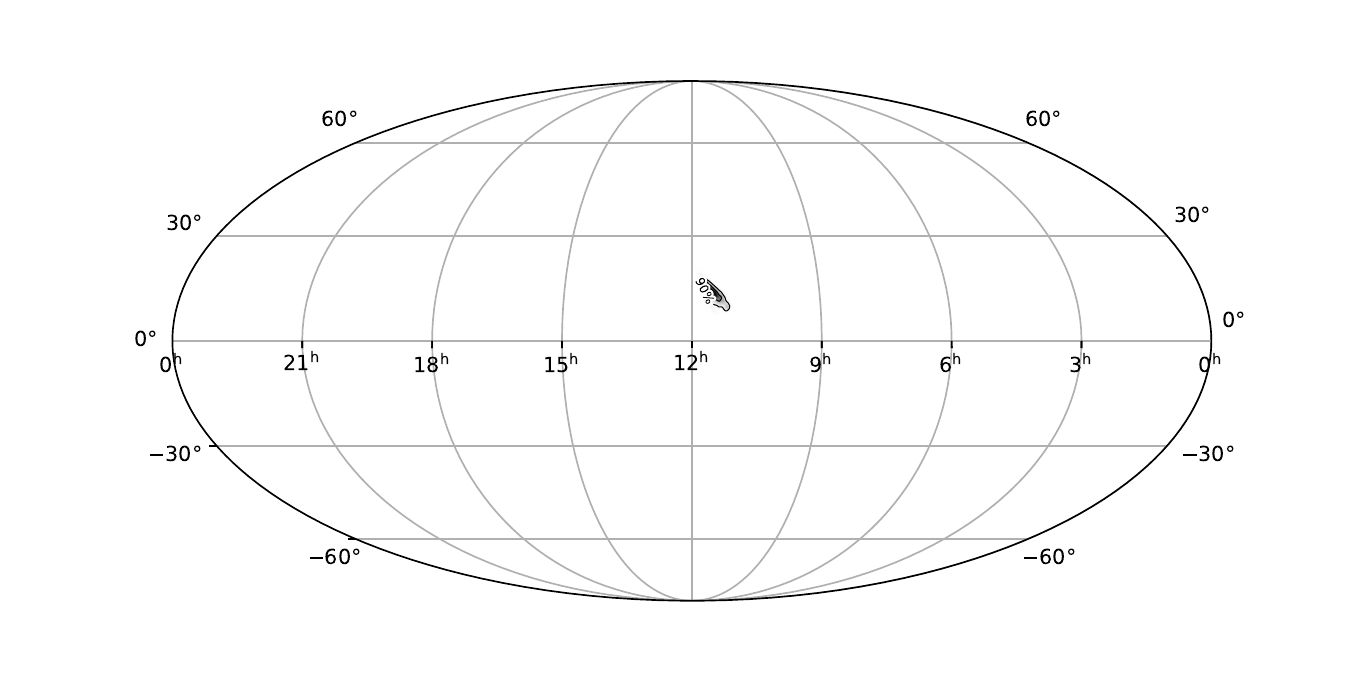}};       
        \node[anchor=south west, draw=black, fill=white, inner sep=1pt] (zoom)
            at (0.01\linewidth, 0.35\linewidth)
            {\includegraphics[width=0.65\linewidth,
            trim={0.72in 0.30in 0.7in 0.30in},clip
            ]{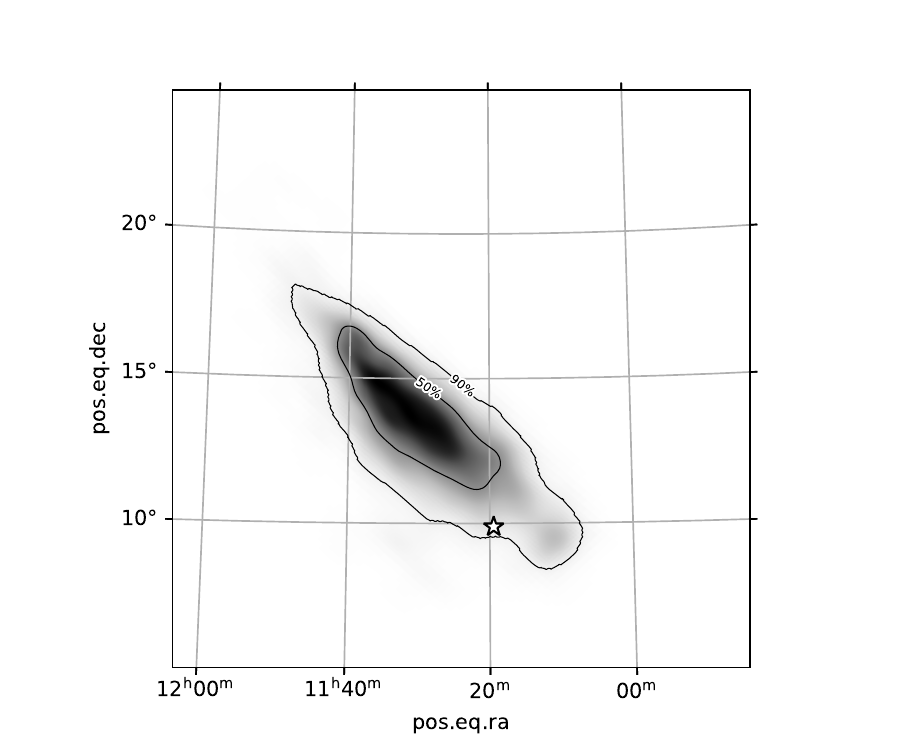}};
        \draw[black, thick]
            (0.516\linewidth, 0.265\linewidth)
            rectangle
            ++(0.035\linewidth, 0.035\linewidth);
        \draw[black]
            (0.515\linewidth, 0.30\linewidth) -- (zoom.south west);
        \draw[black]
            (0.55\linewidth, 0.30\linewidth) -- (zoom.south east);
        \node[anchor=south west,fill=white, inner sep=0pt] (dist)
            at (0.56\linewidth, 0.01\linewidth)
            {\includegraphics[width=0.42\linewidth]{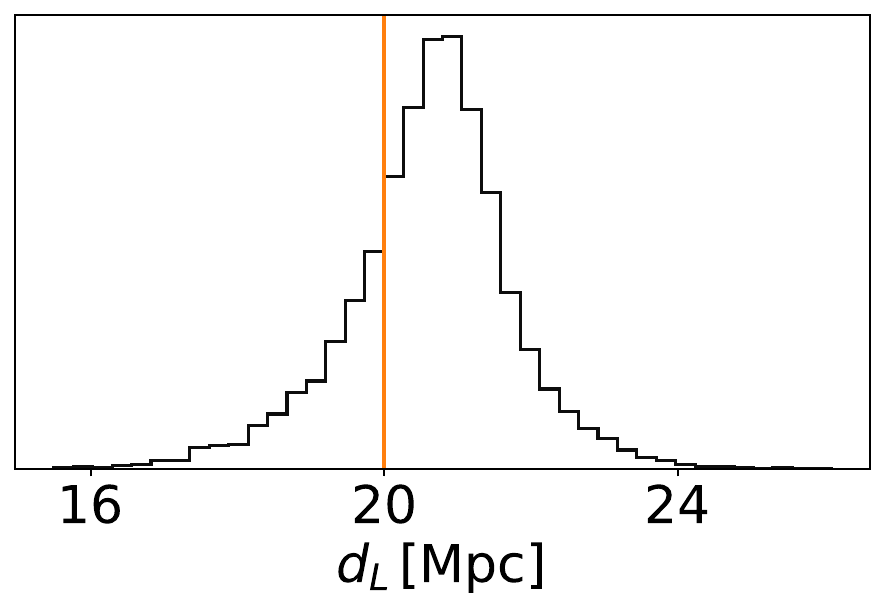}};    
    \end{tikzpicture}
    \caption{
    Sky localization for the recovered signal in a single Cosmic Explorer detector. The main panel shows the full posterior sky map, and the inset provides a magnified view of the highest-posterior-density region. The square in the main panel marks the region shown in the inset. The recovered \(90\%\) credible sky area is approximately \(41\,\mathrm{deg}^2\). The lower-right panel shows the marginalized posterior distribution for the luminosity distance \(d_L\), with the solid orange line indicating the injected value. 
    }
    \label{fig:skymap}
\end{figure}
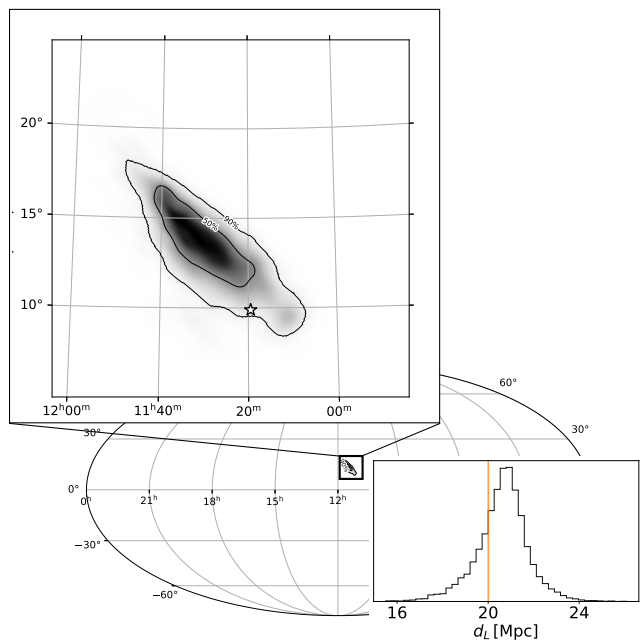

Figure~\ref{fig:skymap} shows the corresponding sky localization and distance, with an inset highlighting the highest-posterior-density region. We find a 90\% credible sky area of approximately $41\,\mathrm{deg}^2$, consistent with the scale found in previous studies~\citep{BNS_localization1,BNS_localization2}. Such localization from a single detector is possible because the signal remains in band for a sufficiently long time that the Earth's rotation modulates the detector response over the course of the observation. This time dependence effectively provides directional information, partially breaking the degeneracy that would otherwise be present in a short-duration single-detector signal, allowing the source position to be constrained far more tightly.

Taken together, these results show that, once the time-dependent detector response is properly modeled, a single high-SNR binary-neutron-star observation in a Cosmic Explorer detector can simultaneously yield precise intrinsic-parameter constraints, sub-kilometer mass--radius measurements, and tight sky localization.

The remaining source parameters are reported in Appendix~\ref{appx:inference}. Further discussion of the parameter degeneracies underlying the observed posterior shifts in Figure~\ref{fig:skymap} is given in Appendix~\ref{app:posterior_shift_diagnostic}.

The reweighting analysis yields a negligible shift in the log evidence, $\Delta \log Z \simeq -0.2$, together with an efficiency close to unity, $\epsilon \simeq 0.992$. This indicates that the ROQ posterior is statistically consistent with the posterior obtained from the standard likelihood. The differences between the weighted and unweighted sample distributions are visually negligible, and we therefore omit the reweighted samples from the figures to avoid additional clutter.

These results indicate that the subbanded ROQ construction remains sufficiently accurate for practical inference even in the presence of a long-duration, time-dependent detector response.

Finally, the inference required approximately $4.2\times10^8$ likelihood evaluations and, using 128 CPU cores, took about 48\,h. Independent timing tests give $\sim 0.023\,\mathrm{s}$ per ROQ likelihood evaluation. Without the ROQ method, we find that each likelihood evaluation is slower by a factor of $\sim 360$, implying that without the ROQ the total runtime is roughly 2\,years on 128 cores, and of order $\mathcal{O}(10^2)$ years on a single CPU.

\section{Conclusions and Discussion}\label{sec:conclusions}

Next-generation gravitational-wave detectors are expected to deliver a substantial improvement in sensitivity. To fully realize their scientific potential, the corresponding data-analysis methods must be able to extract information efficiently and accurately from long-duration, high-SNR signals. Studies such as the present one are therefore essential for demonstrating the practical feasibility of inference in this regime.

In this work, we revise the construction of reduced-order models and reduced-order quadratures for long binary neutron-star signals in the presence of time-dependent detector response. In particular, we address two key limitations of the standard approach: a memory limitation, associated with the large number of frequency grid points and basis elements required at low frequencies, and an accuracy limitation, associated with the degradation introduced during the ROQ construction. We introduce a modified construction strategy based on improved adaptive sampling, disk-backed streaming, and subbanded ROQ construction. Together, these modifications substantially reduce the memory burden, improve numerical stability, and make the construction naturally parallelizable.

Using this framework, we demonstrate Bayesian inference on a binary neutron-star signal in a single Cosmic Explorer detector with SNR exceeding 2000, including time-dependent detector-response effects down to 5\,Hz. To our knowledge, this is the first demonstration of ROQ-based inference in this regime including time-dependent effects. The resulting inference remains accurate under reweighting tests against the standard likelihood, indicating that the construction is sufficiently reliable even in the high-SNR, long-duration limit considered here. Extending the method to a detector network is straightforward, since the total likelihood is just the product of the individual detector likelihoods. The same ROQ framework can therefore be applied to each detector and the resulting likelihood contributions combined in the usual way.

A further consequence of long-duration signals is the possibility of significantly improved sky localization, even with a single detector. As the Earth rotates during the observation, the detector response evolves in time and thereby encodes additional directional information. This is reflected in the present analysis by the small 90\% credible sky area that we obtain. Such accurate localization is particularly relevant for multimessenger astronomy, where it potentially facilitates the identification of electromagnetic counterparts and host galaxies~\citep{em_followup}, and for standard-siren cosmology, where reliable host association enables measurements of the Hubble constant~\citep{H0}, while statistical association with galaxy catalogs can be used for dark-siren cosmological inference~\citep{dark_sirens}.

The masses of the double pulsar PSR~J0737$-$3039A/B are among the most precisely measured neutron-star masses known, with uncertainties of order \(\mathcal{O}(10^{-5})\,M_\odot\)~\citep{double_pulsar}. In our analysis, the chirp mass is recovered with an uncertainty of order \(\mathcal{O}(10^{-6})\,M_\odot\), whereas the inferred component masses have uncertainties of order \(\sim 4\times10^{-3}\,M_\odot\). This shows that even with a single detector, long-duration, high-SNR gravitational-wave observations can determine the chirp mass with a precision comparable to that of the best radio-timing mass measurements, even though the individual component masses remain less tightly constrained.

The inferred equation-of-state constraints provide a further illustration of the scientific reach of such high-SNR observations. From the posterior samples, we infer a polytropic equation of state and the corresponding mass--radius relation, finding that the joint 90\% credible band in the mass–radius plane has a width of
\(\Delta R^{90}\lesssim 1\,\mathrm{km}\) for \(M \gtrsim 1.3\,M_\odot\). Such tight constraints are broadly in line with earlier forecasts for third-generation detectors~\citep{NS_eos_next_gen}. By comparison, current observational radius constraints are typically at the level of \(\sim 1\) km or larger; for example, the updated NICER analysis of PSR~J0740+6620 reports \(R=12.92^{+2.09}_{-1.13}\,\mathrm{km}\) at 68\% credibility~\citep{NS_radi}. While our result is based on an idealized high-SNR signal, it highlights the potential of next-generation gravitational-wave detectors to deliver highly precise radius constraints.

The present study may also have implications beyond next-generation detectors. Even for current-generation instruments, moderately long signals with durations of order 16--32\,s, or longer, may benefit more from careful control of the number of frequency grid points used in the ROQ construction than from enforcing extremely small construction errors. In such cases, one can construct ROQs that are sufficiently accurate for low-to-moderate SNRs while remaining extremely cheap to evaluate. More generally, our results show that the frequency grid design can be used to tune the ROQ construction to the signal regime of interest, trading unnecessary accuracy for substantial computational savings.

\section*{Acknowledgments}
The authors would like to thank Teagan Clarke for her useful suggestions on an earlier version of this manuscript and the anonymous referee for their constructive comments.
This work is supported through the Australian Research Council (ARC) Centre of Excellence CE230100016, Discovery Projects DP220101610 and DP230103088, and LIEF Project LE210100002.
This material is based upon work supported by NSF's LIGO Laboratory which is a major facility fully funded by the National Science Foundation.
The authors are grateful for computational resources provided by the LIGO Laboratory and supported by National Science Foundation Grants PHY-0757058 and PHY-0823459.

We also acknowledge CPU time on OzSTAR,
funded by Swinburne University and the Australian Government. 

LIGO Laboratory and Advanced LIGO are funded by the United States National Science Foundation (NSF) as well as the Science and Technology Facilities Council (STFC) of the United Kingdom, the Max-Planck-Society (MPS), and the State of Niedersachsen/Germany in support of the construction of Advanced LIGO and construction and operation of the GEO600 detector. Additional support for Advanced LIGO was provided by the Australian Research Council. Virgo is funded, through the European Gravitational Observatory (EGO), by the French Centre National de Recherche Scientifique (CNRS), the Italian Istituto Nazionale di Fisica Nucleare (INFN) and the Dutch Nikhef, with contributions by institutions from Belgium, Germany, Greece, Hungary, Ireland, Japan, Monaco, Poland, Portugal, Spain. The construction and operation of KAGRA are funded by Ministry of Education, Culture, Sports, Science and Technology (MEXT), and Japan Society for the Promotion of Science (JSPS), National Research Foundation (NRF) and Ministry of Science and ICT (MSIT) in Korea, Academia Sinica (AS) and the Ministry of Science and Technology (MoST) in Taiwan.

\appendix

\section{\texttt{gwRombusX}}\label{appx:rombus}

\texttt{Rombus} was originally developed as a general framework for reduced-order modeling and is publicly available at~\citet{git_smith}. Subsequently, \texttt{gwRombus} extended this framework with functionality specific to gravitational-wave applications~\citep{git_makai}. Our implementation, \texttt{gwRombusX}, builds on both repositories, with substantial modifications introduced to address the memory and accuracy limitations discussed in the main text.

At a high level, \texttt{gwRombusX} workflow used in this work proceeds as follows:
\begin{enumerate}
    \item Generate a training set of waveforms over the parameter region of interest using the adaptive frequency-sampling strategy described in the main text (see Section~\ref{sec:roq_method}).
    \item Construct a reduced basis on the adaptively sampled grid using a greedy algorithm. Iteratively refine the basis by identifying poorly represented waveforms, and adding them into the training set.
    \item Reconstruct the selected basis waveforms on the full uniform frequency grid and re-orthonormalize them using a Gram--Schmidt procedure.
    \item Apply the empirical interpolation method to determine the interpolation nodes and thereby construct the reduced-order model. This is done iteratively: the first node is chosen as the frequency at which the first basis element has the largest magnitude. Each subsequent node is then chosen from the residual obtained by approximating the next basis element using the nodes already selected, with the new node placed at the frequency where this residual is maximal.
    \item Build the ROQ weights for each frequency subband and combine them into the final likelihood rule.
    \item Validate the resulting ROQ through direct likelihood comparisons and, if necessary, refine the sampling and repeat the construction.
\end{enumerate}

The main additions introduced in \texttt{gwRombusX} include an updated implementation of the dynamic sampling, disk-backed storage of the reduced basis, streaming and blocked linear-algebra operations for large arrays, support for constructing ROQs over multiple frequency subbands (See Appendix~\ref{app:disk_streaming}), and an updated implementation of the time-dependent ROQ likelihood in \bilby. These changes allow the construction to remain feasible for significantly longer and more complex signals than was previously practical.

The implementation used in this work, together with example configuration files and execution scripts, can be found at \href{https://github.com/NirGutt/gwRombusX}{\texttt{gwRombusX}}\footnote{\url{https://github.com/NirGutt/gwRombusX}}. These examples illustrate the construction of the training set, the reduced basis, the empirical interpolant, and the final ROQ likelihood used in the inference studies presented here.

\subsection{Implementation of disk-backed streaming}
\label{app:disk_streaming}

In \texttt{gwRombusX}, we replace the original
in-memory representation of the training set and reduced basis with a disk-backed
architecture. Rather than storing these objects as dense arrays in memory, they
are written to disk during construction and accessed only as needed, an approach referred to here as streaming. This changes the peak memory requirement from scaling with the total size of the training set
and reduced basis to scaling primarily with the size of the data block
being processed at any given time. The key implementation choices for the different stages of the
ROM and ROQ construction (see Sec.~\ref{subsec:rom}) are detailed below.

The training set is distributed across multiple processes. Each process
independently generates and stores its local portion of the training set on
disk, which is possible because the individual training points can be generated independently.

During the construction of the reduced basis, a designated master process reopens the
training-set files in read-only mode and streams the required data directly
from disk. This avoids transferring large arrays between processes and
eliminates the need to keep the complete training set resident in memory.

In addition, a Gram--Schmidt orthogonalization algorithm is implemented to operate on streamed data. Previously constructed basis vectors are read in blocks, and each block is used to compute the projection of the candidate vector onto  the current block of basis vectors. This projection is then subtracted from the candidate vector before the next block is processed. Once all blocks have been considered, the residual is normalized and added to the basis. The number of basis vectors included in each block is determined from a user-specified memory budget together with the data dimensions. This allows the memory usage to be adjusted to systems with different available capacities without changing the underlying algorithm.

The empirical-interpolation stage follows the same streaming approach. The completed reduced basis is accessed directly from disk, and only
the data required at each step are loaded into memory. The interpolation
quantities are likewise constructed in blocks and written incrementally to disk.
This prevents the full reduced basis and interpolation data from being held in
memory simultaneously, allowing the empirical-interpolation stage to retain a
bounded memory footprint.

The main cost of this approach is increased disk input/output compared with
an entirely memory-resident implementation. Smaller memory budgets result in
smaller processing blocks and therefore more frequent disk access, while
larger memory budgets reduce the number of streaming operations. The streamed implementation is mathematically equivalent to the corresponding in-memory implementation and introduces no additional approximation.

\section{Optimal frequency-band spacing}
\label{appx:binning_derivation}

At leading post-Newtonian order, the time to merger of a compact binary from frequency $f$ scales as
\begin{equation}
T(f)=\frac{5}{256}\mathcal{M}_c^{-5/3}(\pi f)^{-8/3}\propto f^{-8/3},
\end{equation}
where $\mathcal{M}_c$ is the chirp mass. Consider a decomposition of the full frequency interval $[f_{\min},f_{\max}]$ into a sequence of bands with edges $\{f_i\}$. The time spent in the $i$th band, bounded by $[f_i,f_{i+1}]$, is
\begin{equation}
\Delta t_i = T(f_i)-T(f_{i+1}).
\end{equation}
Following the dynamic-sampling prescription, the frequency resolution in that band scales as
\begin{equation}
\Delta f_i \sim \frac{1}{\Delta t_i}.
\end{equation}
The corresponding number of frequency bins in the band is therefore
\begin{equation}
n_i \sim \frac{f_{i+1}-f_i}{\Delta f_i}
      \sim (f_{i+1}-f_i)\,[T(f_i)-T(f_{i+1})].
\end{equation}

The total number of bins is obtained by summing over all bands. To determine the optimal band placement, it is convenient to pass to a continuum description. We introduce a continuous band-index variable $x\in[0,1]$, where $x=0$ corresponds to $f_{\min}$ and $x=1$ corresponds to $f_{\max}$. In practice, for $N_{\rm band}$ bands, one may think of $x=i/N_{\rm band}$. The band edges are then described by a monotonic function $f(x)$, and a small interval in $x$ corresponds to one band. Defining \(f'(x)\equiv df/dx\) and \(T'(f)\equiv dT/df\), the band width becomes
\begin{equation}
f_{i+1}-f_i \rightarrow \frac{df}{dx}\,dx = f'(x)\,dx,
\end{equation}
while the corresponding time spent in the band is
\begin{equation}
T(f_i)-T(f_{i+1}) \rightarrow -\frac{dT(f)}{df}\frac{df}{dx}\,dx = -T'(f)\,f'(x)\,dx.
\end{equation}
The contribution of one infinitesimal band therefore becomes
\begin{equation}
dn \propto -T'(f)\,[f'(x)]^2\,dx.
\end{equation}
Since $T(f)\propto f^{-8/3}$, it follows that
\begin{equation}
-T'(f)\propto f^{-11/3},
\end{equation}
and hence minimizing the total number of bins is equivalent, up to an overall constant factor, to minimizing the functional
\begin{equation}
J[f]=\int_0^1 f^{-11/3}(f')^2\,dx.
\end{equation}

The associated Euler--Lagrange equation is
\begin{equation}
\frac{d}{dx}\left(2f^{-11/3}f'\right)
-\left(-\frac{11}{3}f^{-14/3}(f')^2\right)=0,
\end{equation}
which simplifies to
\begin{equation}
\frac{11}{3}(f')^2-2ff''=0.
\end{equation}
This equation can be solved by introducing the substitution
\begin{equation}
u=f^{-5/6}.
\end{equation}
The differential equation then reduces to
\begin{equation}
u''=0,
\end{equation}
whose solution is linear in $x$,
\begin{equation}
u(x)=x(u_{\max}-u_{\min})+u_{\min},
\end{equation}
where
\begin{equation}
u_{\min}=f_{\min}^{-5/6},
\qquad
u_{\max}=f_{\max}^{-5/6}.
\end{equation}
Transforming back to $f$, we obtain
\begin{equation}
f(x)=\left[x\left(f_{\max}^{-5/6}-f_{\min}^{-5/6}\right)+f_{\min}^{-5/6}\right]^{-6/5}.
\end{equation}

We therefore conclude that the optimal band edges are uniformly spaced in the variable $f^{-5/6}$. Equivalently, if the full frequency range is divided into $N_{\rm band}$ subbands, the optimal boundaries are obtained by choosing equal steps in $f^{-5/6}$ between $f_{\min}$ and $f_{\max}$.

\section{Scaling of log-likelihood fluctuations}\label{app:logL_error}

For data $d=h+n$, with signal $h$ and stationary Gaussian noise $n$, the log-likelihood can be written, up to an additive constant, as
\begin{equation}
\log \mathcal{L} = \langle d,h\rangle - \frac{1}{2}\langle h,h\rangle.
\end{equation}
Substituting $d=h+n$ gives
\begin{equation}
\log \mathcal{L} = \frac{1}{2}\langle h,h\rangle + \langle n,h\rangle.
\end{equation}
Defining the optimal SNR by
\begin{equation}
\rho_{\rm opt}^2 \equiv \langle h,h\rangle,
\end{equation}
this becomes
\begin{equation}
\log \mathcal{L} = \frac{1}{2}\rho_{\rm opt}^2 + \langle n,h\rangle.
\end{equation}
Defining the normalized waveform $\hat h = h/\rho_{\rm opt}$, so that $\langle \hat h,\hat h\rangle=1$. Then
\begin{equation}
\langle n,h\rangle = \rho_{\rm opt}\langle n,\hat h\rangle.
\end{equation}
A standard result from matched-filtering theory is that the projection of the noise onto any fixed normalized template is itself a Gaussian random variable, assuming Gaussian detector noise~\citep{product_noise_signal}. Therefore,
\begin{equation}
\langle n,\hat h\rangle \sim \mathcal{N}(0,1),
\end{equation}
and hence
\begin{equation}
\log \mathcal{L} \sim \mathcal{N}\!\left(\frac{1}{2}\rho_{\rm opt}^2,\rho_{\rm opt}^2\right),
\end{equation}
up to an additive constant. Thus, the mean of $\log \mathcal{L}$ scales as $\rho_{\rm opt}^2$, while its fluctuations scale as $\rho_{\rm opt}$.

\section{Optimal distribution of the ROQ error budget}
\label{appx:error_budget}
Starting from the cost model in Eq.~\ref{eq:rom_cost}, we derive the optimal allocation of the total ROQ error budget across subbands. The total error budget is constrained by
\begin{equation}
\sum_i \epsilon_i^2 = \epsilon_{\rm tot}^2.
\end{equation}
To minimize the total cost under this constraint, we introduce a Lagrange multiplier $\lambda$ and define
\begin{equation}
\mathcal{F}(\{\epsilon_i\},\lambda)
=
\sum_i \frac{N_i^{\alpha}\rho_i^{\beta}}{\epsilon_i^{\gamma}}
+\lambda\left(\sum_i \epsilon_i^2-\epsilon_{\rm tot}^2\right).
\end{equation}
Taking the derivative with respect to $\epsilon_i$ gives
\begin{equation}
\frac{\partial \mathcal{F}}{\partial \epsilon_i}
=
-\gamma \frac{N_i^{\alpha}\rho_i^{\beta}}{\epsilon_i^{\gamma+1}}
+2\lambda \epsilon_i
=0.
\end{equation}
Rearranging, we obtain
\begin{equation}
\epsilon_i^{\gamma+2}
=
\frac{\gamma}{2\lambda}\,N_i^{\alpha}\rho_i^{\beta},
\end{equation}
which implies
\begin{equation}\label{eq:d6}
\epsilon_i = A\left(N_i^{\alpha}\rho_i^{\beta}\right)^{1/(\gamma+2)},
\end{equation}
where the constant $A$ is determined by the total-error constraint. Substituting Eq.~\ref{eq:d6} into this constraint gives
\begin{equation}
A^2
\sum_i
\left(N_i^{\alpha}\rho_i^{\beta}\right)^{2/(\gamma+2)}
=
\epsilon_{\rm tot}^2.
\end{equation}
Hence,
\begin{equation}
A
=
\frac{\epsilon_{\rm tot}}
{\sqrt{
\sum_i
\left(N_i^{\alpha}\rho_i^{\beta}\right)^{2/(\gamma+2)}
}}.
\end{equation}
Therefore, the optimal allocation is
\begin{equation}
\epsilon_i
=
\epsilon_{\rm tot}
\frac{
\left(N_i^{\alpha}\rho_i^{\beta}\right)^{1/(\gamma+2)}
}{
\sqrt{
\sum_j
\left(N_j^{\alpha}\rho_j^{\beta}\right)^{2/(\gamma+2)}
}
}.
\end{equation}
For the simple choice $\alpha=\beta=\gamma=1$, this reduces to
\begin{equation}
\epsilon_i \propto (N_i\rho_i)^{1/3},
\end{equation}
which is the expression used in the main text.

\section{Long-duration binary neutron-star inference}\label{appx:inference}

Table~\ref{tab:subband_roq_summary} summarizes the ROQ construction by frequency subband, including the enhancement factor, number of basis elements, local SNRs, and assigned accuracy targets.

At the lowest frequencies, we find it useful to supplement the basic inspiral-based banding of Section~\ref{subsec:accuracy_limitation} with two additional refinements. First, to resolve the time dependence of the detector response due to the Earth's rotation, we require that adjacent frequency samples correspond to no more than \(5\,\mathrm{s}\) of chirp-time evolution. Second, we impose a local chirp-time refinement: if \(\tau(f)\) denotes the time to merger at frequency \(f\), we define \(\Delta \tau_{\rm ref}(f)=\tau(f)-\tau(f+1\,\mathrm{Hz})\) and require the chirp-time separation between adjacent frequency samples to be only 1\% of this reference value. These refinements are applied below \(15\,\mathrm{Hz}\), where the chirp time varies most rapidly and Earth-rotation effects accumulate most strongly. In this sense, they play the same role as the enhancement factors introduced in Section~\ref{subsec:accuracy_limitation}. We adopted them to be deliberately conservative in the region known to be most sensitive to time-dependent effects and ROQ accuracy degradation~\citep{Baker2025}. In practice, this corresponds roughly to an effective enhancement factor of \(N\approx 7\).

\begin{table*}
\centering
\caption{
Summary of the ROQ construction by frequency subband. The first column gives the frequency interval of each subband. The second column lists the enhancement factor $N$ used in the adaptive-sampling prescription of Eq.~\ref{eq:enhancement_factor}. The third column gives the number of basis elements, $N_{\rm basis}$, required in that band. The fourth column shows the SNR accumulated within the corresponding subband. The fifth column reports the target relative accuracy assigned to that band. These target accuracies are computed assuming a total SNR of $\rho=2500$, with the complexity of each subband taken to be proportional to the number of frequency bins in the corresponding full-frequency analysis. The sixth column reports the measured accuracy, quantified using the 95th-percentile criterion described in Section~\ref{sec:roq_accuracy}. The final row shows the corresponding totals over the full frequency range. For the ranges 5--10\,Hz and 10--30\,Hz, the standard enhancement-factor prescription was supplemented by the low-frequency refinement described in the text; the quoted values \(N\sim 7\) indicate the approximate effective enhancement corresponding to that modified sampling.
}
\label{tab:subband_roq_summary}
\begin{tabular}{lccccc}
\hline
Band [Hz] & Enhancement factor $N$ & $N_{\rm basis}$ & Local SNR ($\rho_{i}$)& Target accuracy ($\epsilon_i$)  & Measured accuracy ($r_{95,i}$) \\
\hline
5--10      & $\sim 7$ & 1936    & 331   & $2\times10^{-8}$   & $\mathcal{O}(10^{-10})$ \\
10--30     & $\sim 7$ & 3728    & 1564  & $5\times10^{-8}$   & $\mathcal{O}(10^{-10})$ \\
30--100    & 4 & 704     & 1227  & $8\times10^{-8}$   & $\mathcal{O}(10^{-10})$ \\
100--200   & 5 & 2640    & 474   & $6\times10^{-8}$  & $2.6\times10^{-8}$ \\
200--400   & 2 & 128     & 258   & $7\times10^{-8}$   & $\mathcal{O}(10^{-11})$ \\
400--600   & 2 & 96      & 101   & $5\times10^{-8}$   & $\mathcal{O}(10^{-13})$ \\
600--800   & 2 & 96      & 48    & $4\times10^{-8}$   & $\mathcal{O}(10^{-14})$ \\
800--1000  & 2 & 96      & 26    & $3\times10^{-8}$   & $\mathcal{O}(10^{-14})$ \\
1000--1200 & 2 & 112     & 15    & $3\times10^{-8}$   & $\mathcal{O}(10^{-15})$ \\
1200--1400 & 2 & 128     & 9     & $3\times10^{-8}$   & $\mathcal{O}(10^{-15})$ \\
1400--1600 & 2 & 144     & 6     & $2\times10^{-8}$   & $\mathcal{O}(10^{-16})$ \\
1600--1800 & 2 & 141 & 4     & $2\times10^{-8}$   & $\mathcal{O}(10^{-16})$ \\
1800--2048 & 2 & 155 & 2     & $2\times10^{-8}$   & $\mathcal{O}(10^{-16})$ \\
\hline
5--2048    & - & 10103   & $\sim 2090$ & $1.6\times10^{-7}$ & $\sim 3\times10^{-8}$ \\
\hline
\end{tabular}
\end{table*}

Table~\ref{tab:priors_injection} lists the prior distributions and injection values adopted in the analysis, together with the posterior median and corresponding 90\% credible interval for each inferred parameter.
\begin{table*}
\centering
\caption{Prior distributions, injection values, and inferred 90\% credible intervals for the single-detector high-SNR inference study. The parameters are the chirp mass \(\mathcal{M}_c\), mass ratio \(q\), dimensionless spin magnitudes \(a_1\) and \(a_2\), spin tilt angles \(\theta_1\) and \(\theta_2\), relative spin azimuth \(\phi_{12}\), azimuth of the total angular momentum relative to the orbital angular momentum \(\phi_{JL}\), luminosity distance \(d_L\), declination \(\delta\), right ascension \(\alpha\), inclination angle \(\theta_{JN}\), polarization angle \(\psi\), coalescence phase \(\phi_c\), tidal deformabilities \(\Lambda_1\) and \(\Lambda_2\), and geocenter coalescence time \(t_c\). Angular parameters are given in radians. For non-periodic parameters, the posterior summary is reported as the median together with the 90\% credible interval. For periodic parameters, where the posterior may wrap around the boundary of the coordinate range, the result is instead written as an explicit bracketed interval \([a,b]\), denoting the smallest 90\% credible range on the circle. The \texttt{UniformSourceFrame} prior on \(d_L\) is the source-frame distance prior implemented in \bilby, corresponding to a distribution uniform in source-frame comoving volume~\citep{bilby_paper}.
}
\label{tab:priors_injection}
\begin{tabular}{llll}
\hline
Parameter & Prior & Injection & Posterior (90\% CI) \\
\hline
$\mathcal{M}_c [M_\odot]$ & Uniform [1.1800, 1.1807] & 1.18026510 & $1.18026470_{-0.00000050}^{+0.00000046}$ \\
$q$ & Uniform [0.5, 1.0] & 0.9724 & $0.9725_{-0.0027}^{+0.0028}$ \\
$a_1$ & Uniform [0.0, 0.05] & 0.036 & $0.0416_{-0.0122}^{+0.0072}$ \\
$a_2$ & Uniform [0.0, 0.05] & 0.041 & $0.035_{-0.014}^{+0.013}$ \\
$\theta_1$ & Sine [0.0, 3.14] & 1.98 & $2.07_{-0.30}^{+0.50}$ \\
$\theta_2$ & Sine [0.0, 3.14] & 2.18 & $2.18_{-0.48}^{+0.49}$ \\
$\phi_{12}$ & Uniform [0.0, 6.3] & 5.8 & $0.6\;[4.1,\,3.4]$ \\
$\phi_{JL}$ & Uniform [0.0, 6.3] & 0.1 & $0.0\;[6.1,\,0.2]$ \\
$d_L\,[\mathrm{Mpc}]$ & UniformSourceFrame [1.0, 1000.0]  & 20.0 & $20.7_{-2.1}^{+1.7}$ \\
$\delta$ & Cosine [-1.571, 1.571] & 0.173 & $0.232_{-0.064}^{+0.061}$ \\
$\alpha$ & Uniform [0.0, 6.283] & 2.965 & $3.010_{-0.073}^{+0.059}$ \\
$\theta_{JN}$ & Sine [0.0, 3.142] & 2.016 & $2.066_{-0.071}^{+0.130}$ \\
$\psi$ & Uniform [0.0, 3.142] & 1.678 & $0.20\;[3.13,\,1.76]$ \\
$\phi_c$ & Uniform [0.0, 6.3] & 2.8 & $0.4\;[4.3,\,2.8]$ \\
$\Lambda_1$ & Uniform [0.0, 1000.0] & 483.2 & $460_{-162}^{+140}$ \\
$\Lambda_2$ & Uniform [0.0, 1000.0] & 771.0 & $795_{-154}^{+178}$ \\
$t_c\,[\mathrm{s}]$ & Uniform [-0.1, 0.1] & 0.0351 & $0.0363_{-0.0012}^{+0.0012}$ \\
\hline
\end{tabular}
\end{table*}

\begin{figure*}
    \centering
    \includegraphics[width=1\linewidth]{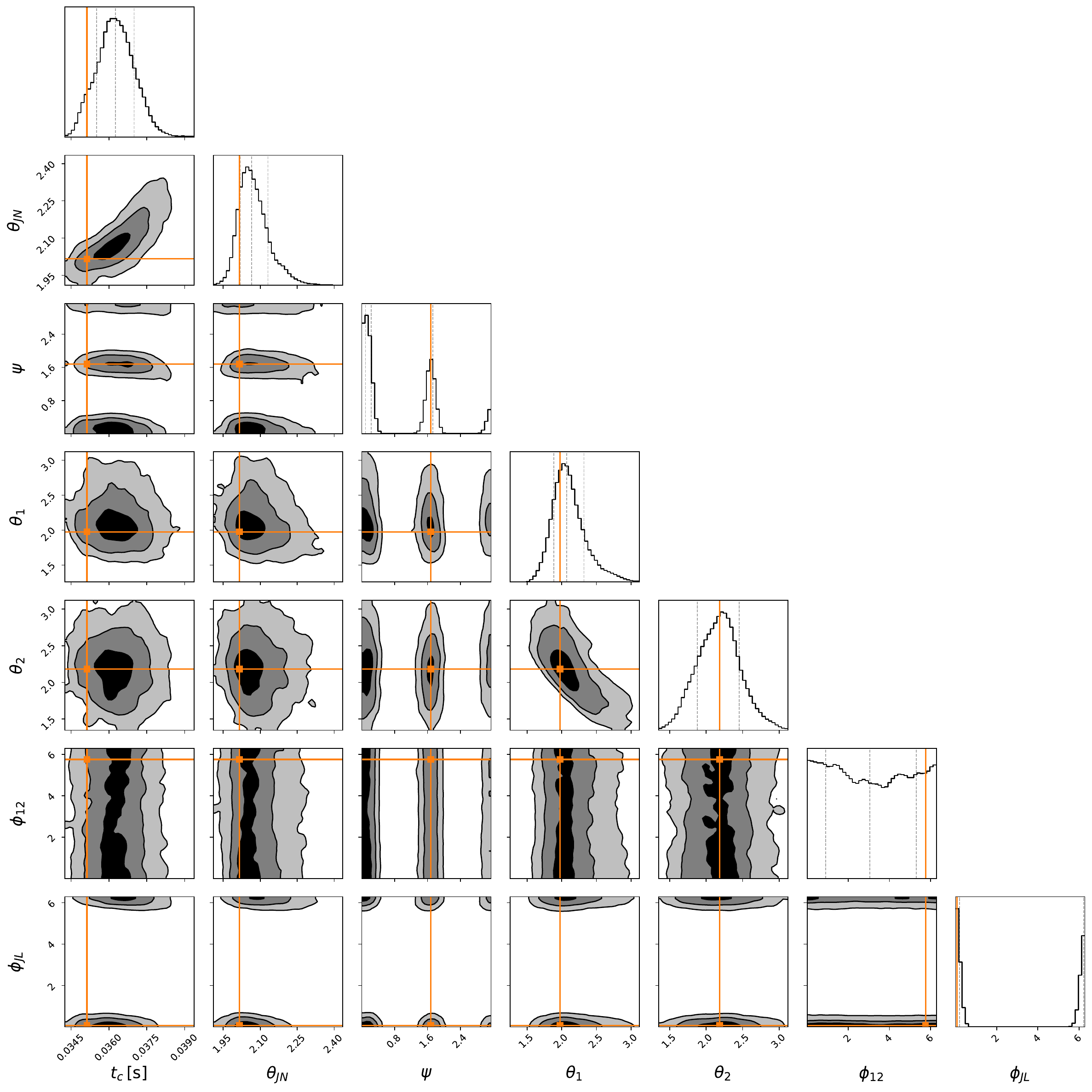}
    \caption{
    Marginalized posterior distributions for the source parameters omitted from the main text. The corner plot shows the time of coalescence $t_c$ as well as the angular parameters \(\theta_1\), \(\theta_2\), \(\phi_{12}\), \(\phi_{JL}\), \(\theta_{JN}\), and \(\psi\). The solid orange lines indicate the injected values, while the dashed lines in the one-dimensional panels show the \(\pm 1\sigma\) credible interval around the posterior median.
    }
    \label{fig:pe_appendix1}
\end{figure*}

\begin{figure*}
    \centering
    \includegraphics[width=0.9\linewidth]{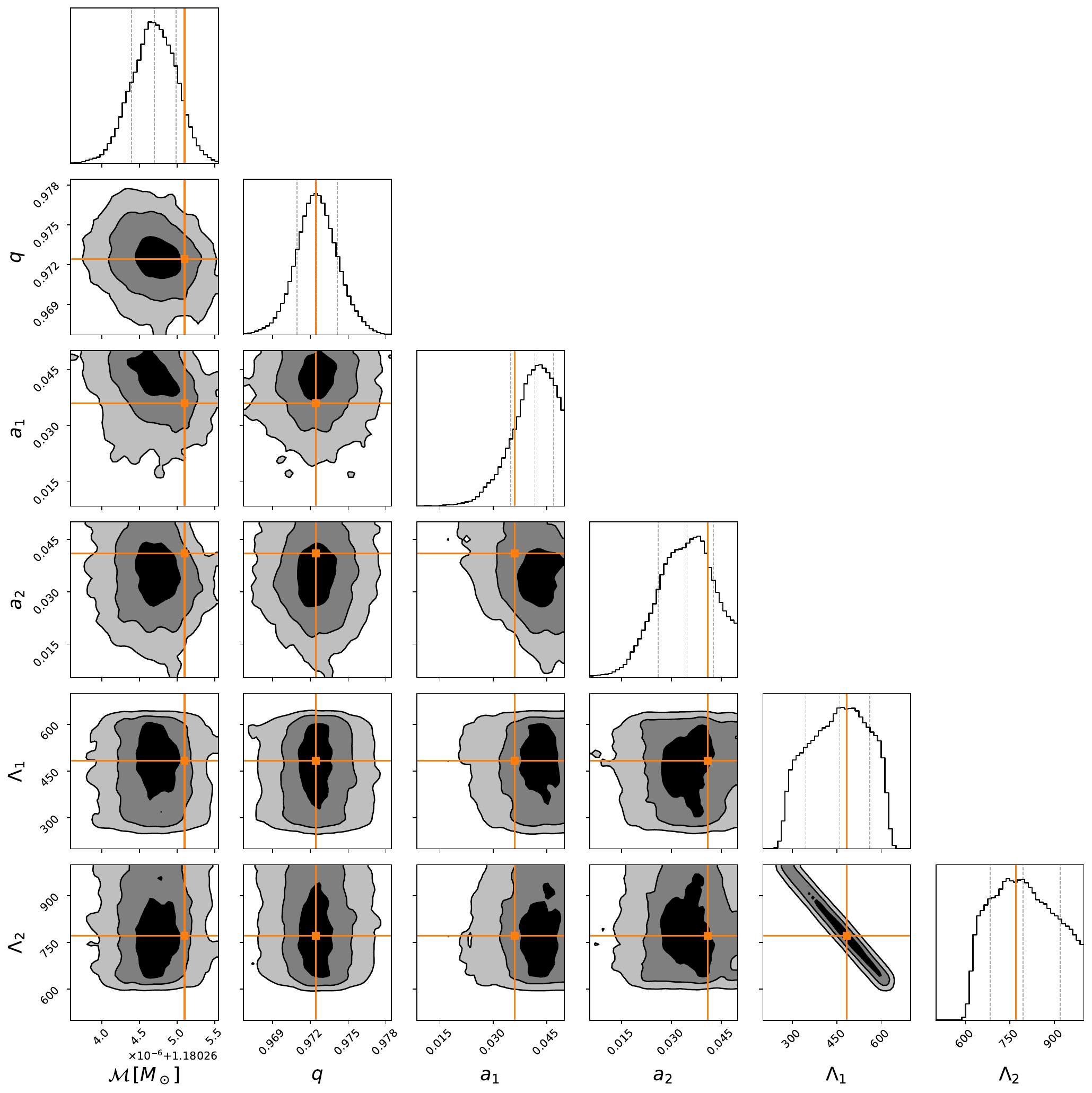}
    \caption{
    Marginalized posterior distributions for the source parameters omitted from the main text. The corner plot shows the source-frame chirp mass \(\mathcal{M}\), mass ratio \(q\), spin magnitudes \(a_1\) and \(a_2\), together with the tidal deformabilities \(\Lambda_1\) and \(\Lambda_2\). The solid orange lines indicate the injected values, while the dashed lines in the one-dimensional panels show the \(\pm 1\sigma\) credible interval around the posterior median.
    }
    \label{fig:pe_appendix2}
\end{figure*}

Figures~\ref{fig:pe_appendix1} and~\ref{fig:pe_appendix2} show the marginalized posterior distributions for the source parameters not displayed in the main text. These results complement the representative posteriors shown in Section~\ref{sec:infernce_results}.

\section{Restricted-inference tests of the posterior shift}
\label{app:posterior_shift_diagnostic}

In the zero-noise inference run presented in Section~\ref{sec:infernce}, one might expect the posterior density to be centered near the injected parameter values. In practice, the posterior distribution peak shifts slightly away from the injection, although the injected values remain inside the high-posterior-density region. To understand this behavior, we perform a sequence of restricted inference runs in which selected parameters are fixed to their injected values.

We first progressively fix subsets of parameters to identify which degrees of freedom are responsible for the shift. We find that when fixing the spin-orientation angles, $\phi_{JL}$, $\phi_{12}$, $\theta_1$, $\theta_2$ and the spin magnitudes $a_1$ and $a_2$, the posterior recenters on the injection.

\begin{figure*}
    \centering
    \includegraphics[width=0.49\linewidth]{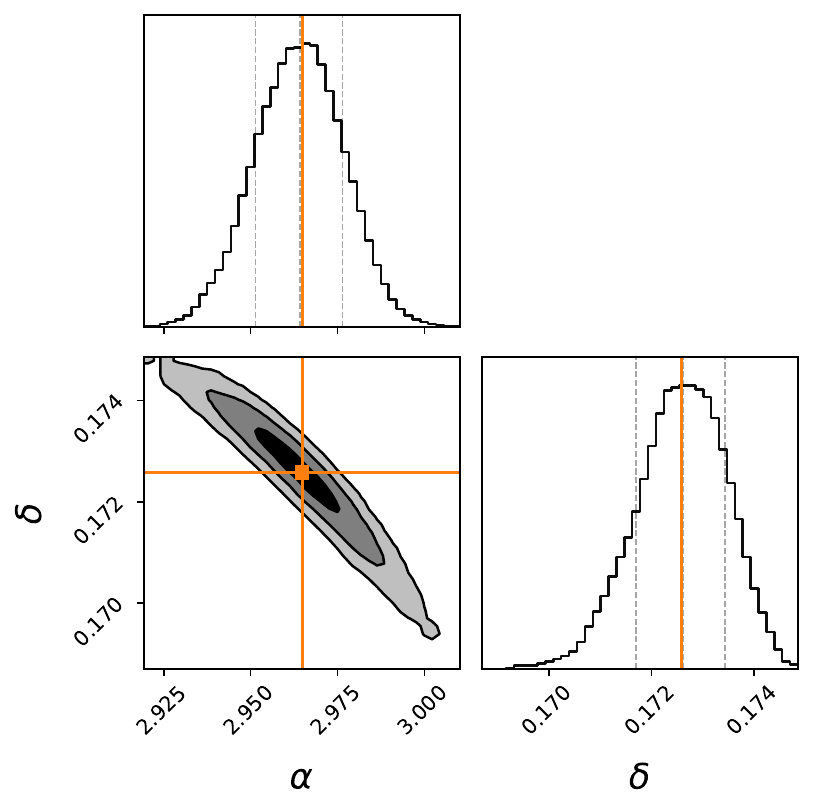}
    \includegraphics[width=0.49\linewidth]{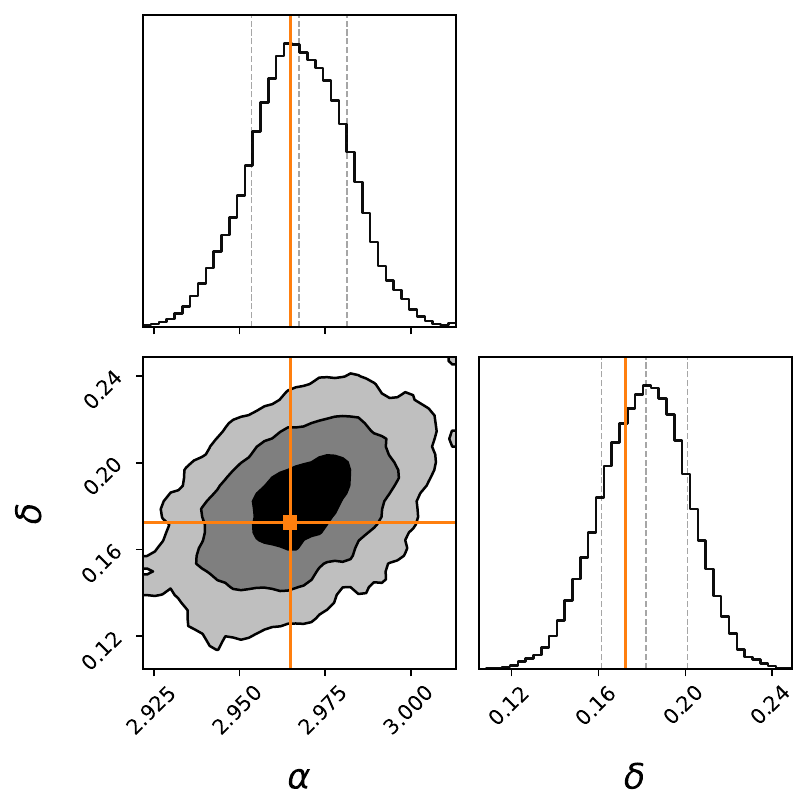}
    \caption{
    Restricted-inference tests for the shifted posterior peaks. 
    Left: sky location posterior when sampling $(\alpha,\delta,\theta_{JN},a_1,a_2)$ with $t_c$ fixed.
    Right: sky location posterior when $t_c$ is added to the sampled parameters.
    The injected values are marked by orange lines.
    The reappearance of the displaced sky posterior when $t_c$ is varied indicates a combined time--sky--inclination--spin degeneracy.
    }
    \label{fig:restricted_recovery}
\end{figure*}

To further isolate the effect, we compare two restricted runs. When sampling only
$\alpha$, $\delta$, $\theta_{JN}$, $a_1$, $a_2$ with $t_c$ fixed, the posterior remains close to the injection. When $t_c$ is added back to the sampled parameters, the displaced correlated structure reappears. 
These runs indicate that the shift is driven by a combined degeneracy involving coalescence time, sky position, inclination, and spin-dependent waveform effects. Representative examples are shown in Fig.~\ref{fig:restricted_recovery}.

As an additional check, we evaluate the likelihood at both the injected parameters and the maximum-likelihood posterior sample. The maximum-likelihood point is only $\Delta\log\mathcal{L}\simeq0.2$ below the injection point, indicating that the injection is not significantly favored. We therefore interpret the displaced peaks as a posterior-volume effect along this degeneracy, rather than as ROQ or sampler failure.

This interpretation is physically plausible for a long-duration signal observed by a single detector. Changes in sky position affect the detector time delay and the Earth-rotation-induced antenna-pattern modulation; changes in $t_c$ alter the time and phase alignment; and changes in inclination and spin parameters can compensate the resulting waveform differences. These correlations create a high-likelihood ridge in which nearby parameter combinations produce nearly indistinguishable detector-frame signals. This highlights the need to use full posterior sampling, as local Gaussian approximations such as the Fisher matrix may fail to capture this extended, correlated, and non-Gaussian high-likelihood structure.

\bibliographystyle{apsrev4-2}
\bibliography{ref}

\end{document}